\pdfoutput=1
\documentclass[prb,twocolumn,showpacs,aps]{revtex4}
\usepackage{graphicx,graphics,color,epsfig}% Include figure files
\usepackage{bm}
\usepackage{amsmath}
\usepackage{amssymb}
\usepackage{epstopdf}

\makeatletter

\newcommand{\Rmnum}[1]{\expandafter\@slowromancap\romannumeral #1@}
\makeatother

\usepackage{color}

\begin{document}

\title{Testing the $d^{*}_{x^{2}-y^{2}}$-wave pairing symmetry by quasiparticle interference in BiS$_{2}$-based superconductors}

\author{Yi Gao,$^{1}$ Tao Zhou,$^{2}$ Huaixiang Huang,$^{3}$ Peiqing Tong,$^{1}$ and Qiang-Hua Wang $^{4}$}
\affiliation{$^{1}$Department of Physics and Institute of Theoretical Physics,
Nanjing Normal University, Nanjing, 210023, China\\
$^{2}$College of Science, Nanjing University of Aeronautics and
Astronautics, Nanjing, 210016, China\\
$^{3}$Department of Physics, Shanghai University, Shanghai, 200444, China\\
$^{4}$National Laboratory of Solid State Microstructures, Nanjing University, Nanjing, 210093, China}

\begin{abstract}
The quasiparticle interference (QPI) patterns in BiS$_{2}$-based superconductors are theoretically investigated by taking into account the spin-orbital coupling and assuming the recently proposed $d^{*}_{x^{2}-y^{2}}$-wave pairing symmetry. We found two distinct scattering wave vectors whose evolution can be explained based on the evolution of the constant-energy contours. The QPI spectra presented in this paper can thus be compared with future scanning tunneling microscopy experiments to test whether the pairing symmetry is $d^{*}_{x^{2}-y^{2}}$-wave in BiS$_{2}$-based superconductors.

\end{abstract}


\maketitle

\emph{Introduction}.---The recent discovery of superconductivity in a new family of BiS$_{2}$-based superconductors \cite{Bi4O4S3} has attracted much attention due to its similarities with the cuprates and iron pnictides. For example, it has a layered crystal structure composed of a stacking of spacer layers and BiS$_{2}$ layers, where, like the CuO and FeAs layers, superconductivity is believed to occur. The initially reported superconducting (SC) transition temperature was $T_{c}=4.5$K in Bi$_{4}$O$_{4}$S$_{3}$ \cite{Bi4O4S3} and later it was found that $Re$O$_{1-x}$F$_{x}$BiS$_{2}$ ($Re=$La, Nd, Ce and Pr) and Sr$_{1-x}$La$_{x}$FBiS$_{2}$ can also exhibit superconductivity.~\cite{LaOFBiS2,NdOFBiS2,CeOFBiS2,PrOFBiS2,SrLaFBiS2} Furthermore, similar to the cuprates and iron pnictides, the parent compounds of the BiS$_{2}$-based superconductors are band insulators (or semiconductors) and superconductivity is induced by electron doping, while $T_{c}$ reaches a maximum at a doping level $x$ at or slightly above $0.5$ for many compounds.~\cite{Bi4O4S3,LaOFBiS2,CeOFBiS2,SrLaFBiS2_pd} Up to now, the highest $T_{c}=10.6$K was reported in LaO$_{0.5}$F$_{0.5}$BiS$_{2}$.~\cite{LaOFBiS2} The above findings suggest that the BiS$_{2}$-based superconductors may also have a relatively high transition temperature and whether or not the SC mechanisms in this kind of materials share some commonalities with the cuprates and iron pnictides needs to be addressed.

At present, the pairing symmetry, which is crucial to understand the SC pairing mechanisms in BiS$_{2}$-based superconductors, is still under debate. Theoretically, band structure calculations show that there exists a strong Fermi-surface (FS) nesting with the nesting wave vector $(k_{x}/\pi,k_{y}/\pi)\approx(1,1)$.~\cite{usui,wan} Thus as widely believed for the iron pnictides, the strong FS nesting, when combined with the electron correlation, could enhance spin or orbital fluctuations and lead to an unconventional pairing symmetry in BiS$_{2}$-based superconductors.~\cite{dagotto,hu} On the other hand, the strong FS nesting was also proposed to enhance the electron-phonon interaction and make the BiS$_{2}$-based superconductors conventional $s$-wave superconductors.~\cite{wan,huang,yildirim} On the experimental side, the measured magnetic penetration depth as a function of temperature can be fitted by assuming a fully gapped conventional $s$-wave pairing symmetry,~\cite{putti,patnaik,Morenzoni} thus supporting the electron-phonon interaction as the pairing glue in BiS$_{2}$-based superconductors. However, later neutron scattering experiments suggested that the electron-phonon coupling in LaO$_{0.5}$F$_{0.5}$BiS$_{2}$ could be much weaker than previously expected.~\cite{lee} Furthermore, recent scanning tunneling microscopy (STM) measurements found a large ratio $2\Delta/k_{B}T_{c}\approx17$ in Bi$_{4}$O$_{4}$S$_{3}$ and NdO$_{1-x}$F$_{x}$BiS$_{2}$,~\cite{wen1,wen2} which far exceeds the value of $3.53$ expected for conventional $s$-wave superconductors, thus indicating an unconventional pairing symmetry in BiS$_{2}$-based superconductors. In addition, a recent theoretical work proposed that there is a dominant triplet component coexisting with the subdominant singlet one in the pairing function, due to the spin-orbital coupling (SOC) resulting from the heavy bismuth atoms. This pairing respects time-reversal symmetry and is $d^{*}_{x^{2}-y^{2}}$-wave with respect to joint spin-lattice operations. In this case, the BiS$_{2}$-based superconductors can be categorized as a time-reversal-invariant weak topological superconductor.~\cite{qhwang}

Previously, several methods have been theoretically proposed to probe the pairing symmetry in BiS$_{2}$-based superconductors.~\cite{zhou,liu,slliu} However, the pairing functions adopted in those works are phenomenological and are not derived based on realistic calculations, while the above mentioned $d^{*}_{x^{2}-y^{2}}$-wave pairing symmetry has not been taken into account. Therefore in this paper, we propose to measure the quasiparticle interference (QPI)~\cite{QPI} as a method to test the $d^{*}_{x^{2}-y^{2}}$-wave pairing symmetry in BiS$_{2}$-based superconductors. The idea is, in realistic materials, if an area of the surface contains dilute randomly distributed impurities, an incoming wave will be scattered into an outgoing wave and the interference between them will lead to a spatial modulation of the local density of states (LDOS), which can be measured by STM.~\cite{hoffman} Furthermore, if the impurities are weak, the evolution of the modulation wave vectors with energy can reflect the energy dependence of the constant-energy contour (CEC), which is determined by the electronic band structure as well as the SC pairing symmetry. Thus by inspecting the modulation wave vectors, the information of the electronic band structure and the SC pairing symmetry can be obtained.

\emph{Method}.---We start with the lattice model proposed in Ref.~\onlinecite{qhwang}, which explicitly takes the SOC into account. The Hamiltonian can be written as
\begin{eqnarray}
\label{h}
H&=&\frac{1}{2}\sum_{\mathbf{k}}\psi_{\mathbf{k}}^{\dag}M_{\mathbf{k}}\psi_{\mathbf{k}},\nonumber\\
\psi_{\mathbf{k}}^{\dag}&=&(c_{\mathbf{k}1\uparrow}^{\dag},c_{\mathbf{k}2\uparrow}^{\dag},c_{\mathbf{k}1\downarrow}^{\dag},c_{\mathbf{k}2\downarrow}^{\dag},
c_{-\mathbf{k}1\uparrow},c_{-\mathbf{k}2\uparrow},c_{-\mathbf{k}1\downarrow},c_{-\mathbf{k}2\downarrow}),\nonumber\\
M_{\mathbf{k}}&=&\begin{pmatrix}
A_{\mathbf{k}}&D_{\mathbf{k}}\\D_{\mathbf{k}}^{\dag}&-A_{-\mathbf{k}}^{T}
\end{pmatrix},\nonumber\\
A_{\mathbf{k}}&=&\chi_{\mathbf{k}}+\xi_{\mathbf{k}},\nonumber\\
\chi_{\mathbf{k}}&=&\begin{pmatrix}
\epsilon_{A\mathbf{k}}-\mu&\epsilon_{xy\mathbf{k}}&0&0\\
\epsilon_{xy\mathbf{k}}&\epsilon_{B\mathbf{k}}-\mu&0&0\\
0&0&\epsilon_{A\mathbf{k}}-\mu&\epsilon_{xy\mathbf{k}}\\
0&0&\epsilon_{xy\mathbf{k}}&\epsilon_{B\mathbf{k}}-\mu
\end{pmatrix},\nonumber\\
\epsilon_{A\mathbf{k}}&=&-t_{1}(\cos k_{x}+\cos k_{y})-t_{2}(\cos k_{x}-\cos k_{y})\nonumber\\
& &+t_{3}\cos k_{x}\cos k_{y}+t_{4}(\cos 2k_{x}-\cos 2k_{y}),\nonumber\\
\epsilon_{B\mathbf{k}}&=&-t_{1}(\cos k_{x}+\cos k_{y})+t_{2}(\cos k_{x}-\cos k_{y})\nonumber\\
& &+t_{3}\cos k_{x}\cos k_{y}-t_{4}(\cos 2k_{x}-\cos 2k_{y}),\nonumber\\
\epsilon_{xy\mathbf{k}}&=&-t_{5}\sin k_{x}\sin k_{y},\nonumber\\
\xi_{\mathbf{k}}&=&-\lambda\sigma_{z}\otimes\tau_{2}-\gamma_{s}(\sin k_{x}\sigma_{y}-\sin k_{y}\sigma_{x})\otimes\tau_{0}\nonumber\\
& &-\gamma_{d}(\sin k_{x}\sigma_{y}+\sin k_{y}\sigma_{x})\otimes\tau_{3},\nonumber\\
D_{\mathbf{k}}&=&\Delta_{0}(g_{\mathbf{k}}+\gamma_{\mathbf{k}}),\nonumber\\
g_{\mathbf{k}}&=&-0.01i\sigma_{y}\otimes\tau_{3},\nonumber\\
\gamma_{\mathbf{k}}&=&0.11(\sin k_{y}\cos k_{x}\sigma_{x}+\sin k_{x}\cos k_{y}\sigma_{y})i\sigma_{y}\otimes\tau_{0}\nonumber\\
& &-0.58(\sin k_{x}\cos k_{y}\sigma_{x}+\sin k_{y}\cos k_{x}\sigma_{y})i\sigma_{y}\otimes\tau_{1}\nonumber\\
& &+0.39(\sin k_{y}\cos k_{x}\sigma_{x}-\sin k_{x}\cos k_{y}\sigma_{y})i\sigma_{y}\otimes\tau_{3},\nonumber\\
\end{eqnarray}
where $c_{\mathbf{k}1\uparrow}^{\dag}$ and $c_{\mathbf{k}2\uparrow}^{\dag}$ create a spin-up electron with momentum $\mathbf{k}$ in the $p_{x}$ and $p_{y}$ orbitals, respectively. Here $t_{1-5}=0.167,0.1,1.948,0.028,1.572$ and $\mu$ is the chemical potential which is adjusted according to the electron filling $x$. $(\lambda,\gamma_{s},\gamma_{d})=(0.5,0.01,0.08)$ are the SOC constants. $D_{\mathbf{k}}$ describes the pairing term of the Hamiltonian at $x=0.55$, the electron doping level we focused on in this paper and we set $\Delta_{0}=0.03$. $\sigma$ and $\tau$ are the Pauli matrices acting on the spin and orbital bases, respectively.

When a single impurity is located at the origin, the impurity Hamiltonian can be written as
\begin{eqnarray}
\label{himp}
H_{imp}&=&V_{s}\sum_{l=1}^{2}\sum_{\sigma=\uparrow,\downarrow}c_{0l\sigma}^{\dag}c_{0l\sigma}\nonumber\\
&=&\frac{V_{s}}{N}\sum_{l=1}^{2}\sum_{\sigma=\uparrow,\downarrow}\sum_{\mathbf{k},\mathbf{k}^{'}}c_{\mathbf{k}l\sigma}^{\dag}c_{\mathbf{k}^{'}l\sigma},
\end{eqnarray}
with $N$ being the system size ($396\times396$ throughout the paper). We consider nonmagnetic impurity only, diagonal in the orbital basis and with a scattering strength $V_{s}$=4$\Delta_{0}$ for definiteness. Following the standard $T$-matrix procedure,~\cite{zhu} the Green's function matrix is defined as
\begin{eqnarray}
\label{gt}
g(\mathbf{k},\mathbf{k}^{'},\tau)=-\langle T_{\tau}\psi_{\mathbf{k}}(\tau)\psi_{\mathbf{k}^{'}}^{\dag}(0)\rangle,
\end{eqnarray}
and
\begin{eqnarray}
\label{gw}
g(\mathbf{k},\mathbf{k}^{'},\omega)&=&\delta_{\mathbf{k}\mathbf{k}^{'}}g_{0}(\mathbf{k},\omega)\nonumber\\
&&+g_{0}(\mathbf{k},\omega)T(\omega)g_{0}(\mathbf{k}^{'},\omega).
\end{eqnarray}
Here $g_{0}(\mathbf{k},\omega)$ is the Green's function in the absence of the impurity and can be written as
\begin{eqnarray}
\label{g0}
g_{0}(\mathbf{k},\omega)&=&[(\omega+i0^{+})I-M_{\mathbf{k}}]^{-1},\nonumber\\
T(\omega)&=&[I-\frac{U}{N}\sum_{\mathbf{q}}g_{0}(\mathbf{q},\omega)]^{-1}\frac{U}{N},\nonumber\\
\end{eqnarray}
where $I$ is a $8\times8$ unit matrix and
\begin{eqnarray}
\label{u}
U_{lm}=\begin{cases}
V_{s}&\text{$l=m=1,\ldots,4$},\\
-V_{s}&\text{$l=m=5,\ldots,8$},\\
0&\text{otherwise}.
\end{cases}
\end{eqnarray}
The experimentally measured LDOS is expressed as
\begin{eqnarray}
\label{rour}
\rho(\mathbf{r},\omega)&=&-\frac{1}{\pi}\sum_{l=1}^{2}\sum_{\sigma=\uparrow,\downarrow}{\rm Im}\langle\langle c_{\mathbf{r}l\sigma}|c_{\mathbf{r}l\sigma}^{\dag}\rangle\rangle_{\omega+i0^{+}}\nonumber\\
&=&-\frac{1}{\pi N}\sum_{m=1}^{4}\sum_{\mathbf{k},\mathbf{k}^{'}}{\rm Im}\Big{[}g_{mm}(\mathbf{k},\mathbf{k}^{'},\omega)e^{-i(\mathbf{k}-\mathbf{k}^{'})\cdot\mathbf{r}}\Big{]},\nonumber\\
\end{eqnarray}
and its Fourier transform is defined as $\rho(\mathbf{q},\omega)=\sum_{\mathbf{r}}\rho(\mathbf{r},\omega)e^{i\mathbf{q}\cdot\mathbf{r}}$, which can be expressed as
\begin{widetext}
\begin{eqnarray}
\label{rouq}
\rho(\mathbf{q},\omega)&=&-\frac{1}{2\pi}\sum_{m=1}^{4}\sum_{\mathbf{k}}{\rm Im}[g_{mm}(\mathbf{k},\mathbf{k}+\mathbf{q},\omega)+g_{mm}(\mathbf{k},\mathbf{k}-\mathbf{q},\omega)]\nonumber\\
&&+i{\rm Re}[g_{mm}(\mathbf{k},\mathbf{k}+\mathbf{q},\omega)-g_{mm}(\mathbf{k},\mathbf{k}-\mathbf{q},\omega)].
\end{eqnarray}
\end{widetext}

\begin{figure}
\includegraphics[width=1\linewidth]{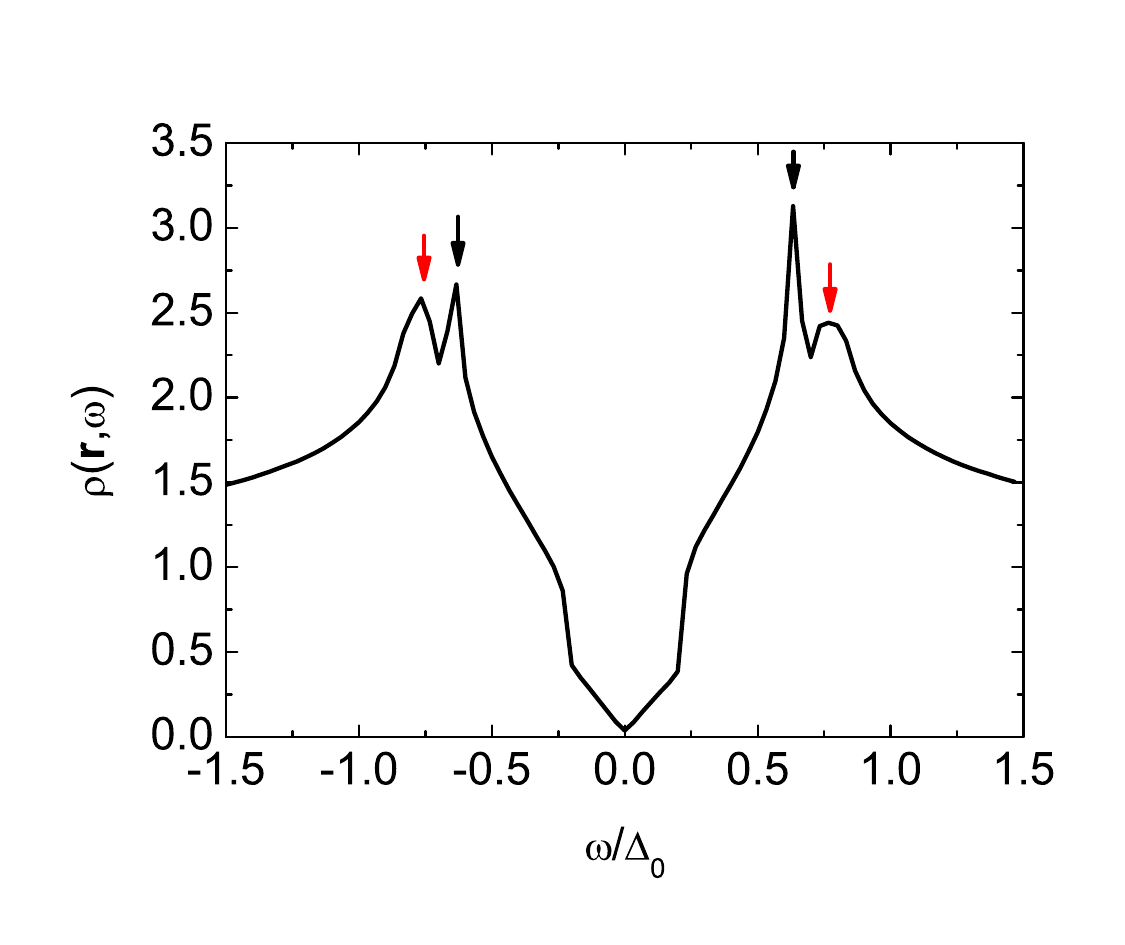}
 \caption{\label{dos} (Color online) $\rho(\mathbf{r},\omega)$ as a function of the reduced energy $\omega/\Delta_{0}$, in the absence of the impurity.}
\end{figure}

\begin{figure*}
\includegraphics[width=0.95\linewidth]{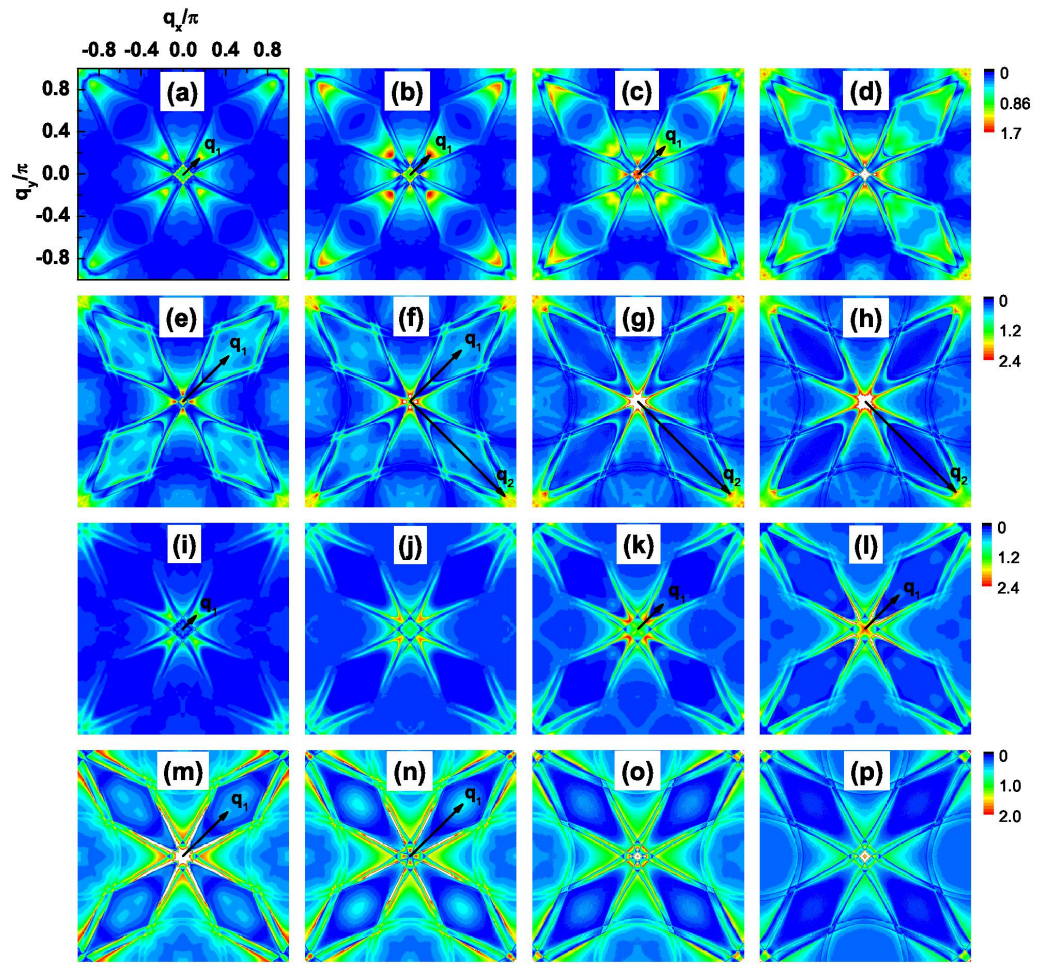}
 \caption{\label{qpi} (Color online) $|\rho(\mathbf{q},\omega)|$ at fixed $\omega$. The point at $\mathbf{q}=0$ is neglected in order to show weaker features at other wave vectors. (a-h) $\omega/\Delta_{0}=-0.2,-0.3,-0.4,-0.5,-0.6,-0.7,-0.8,-0.9$; (i-p) $\omega/\Delta_{0}=0.2,0.3,0.4,0.5,0.6,0.7,0.8,0.9$. Each row shares the same color scale.}
\end{figure*}

\begin{figure}
\includegraphics[width=1\linewidth]{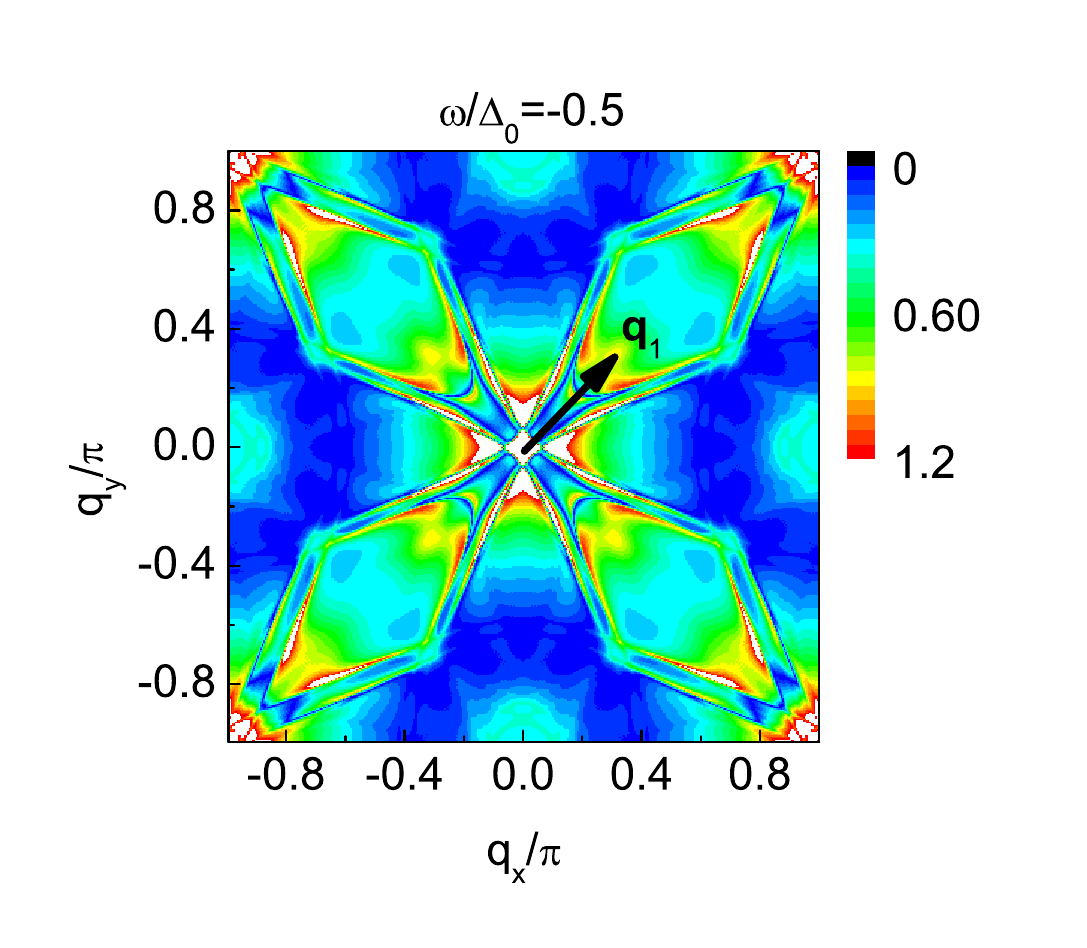}
 \caption{\label{-0.5} (Color online) Same as Fig. \ref{qpi}(d), but with a different color scale.}
\end{figure}

\begin{figure}
\includegraphics[width=1\linewidth]{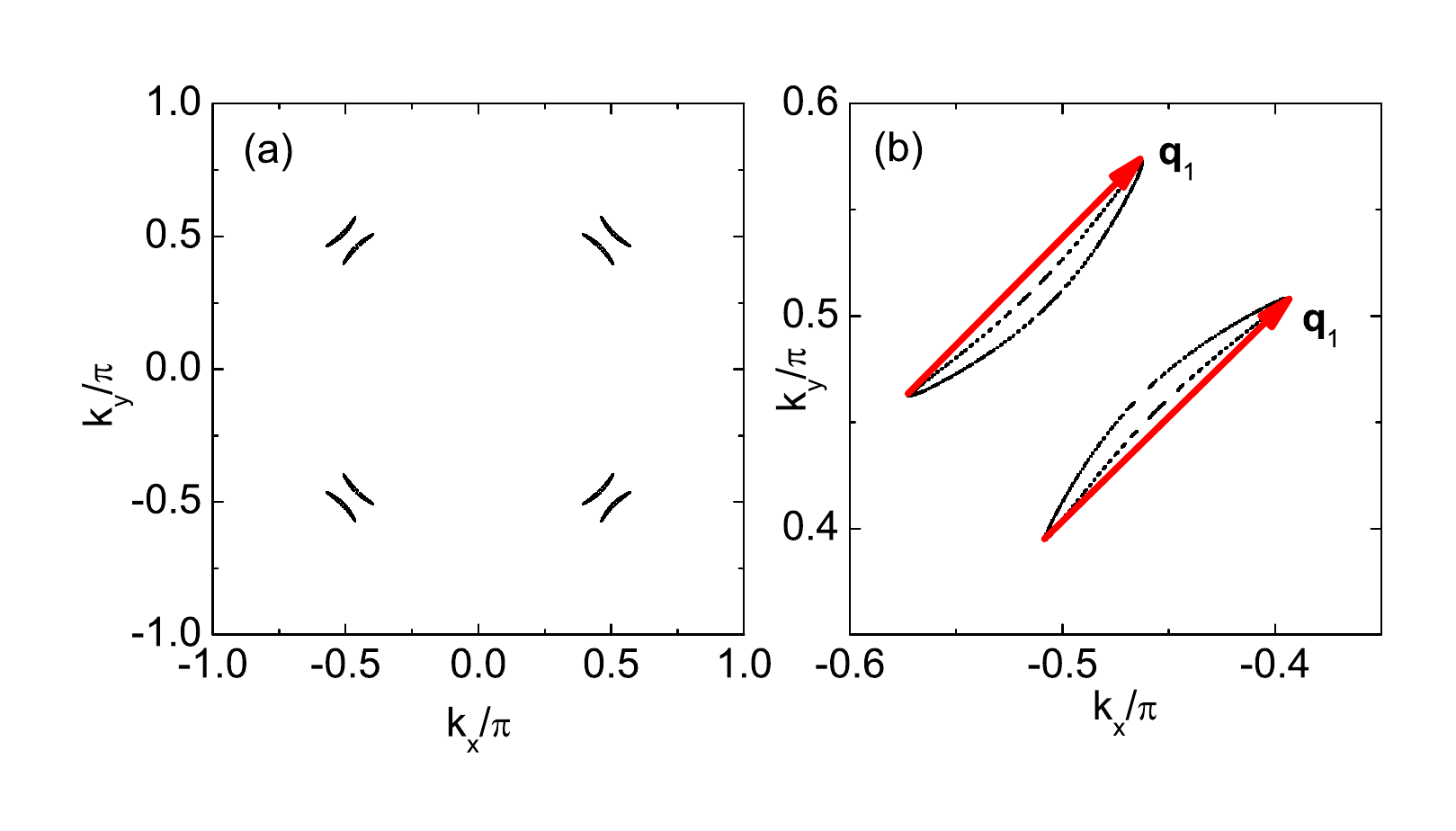}
 \caption{\label{0.2} (Color online) (a) The CEC at $|\omega|/\Delta_{0}=0.2$. (b) shows the CEC around $(k_{x}/\pi,k_{y}/\pi)=(-0.5,0.5)$. The red arrows denote the characteristic scattering wave vectors.}
\end{figure}

\begin{figure}
\includegraphics[width=1\linewidth]{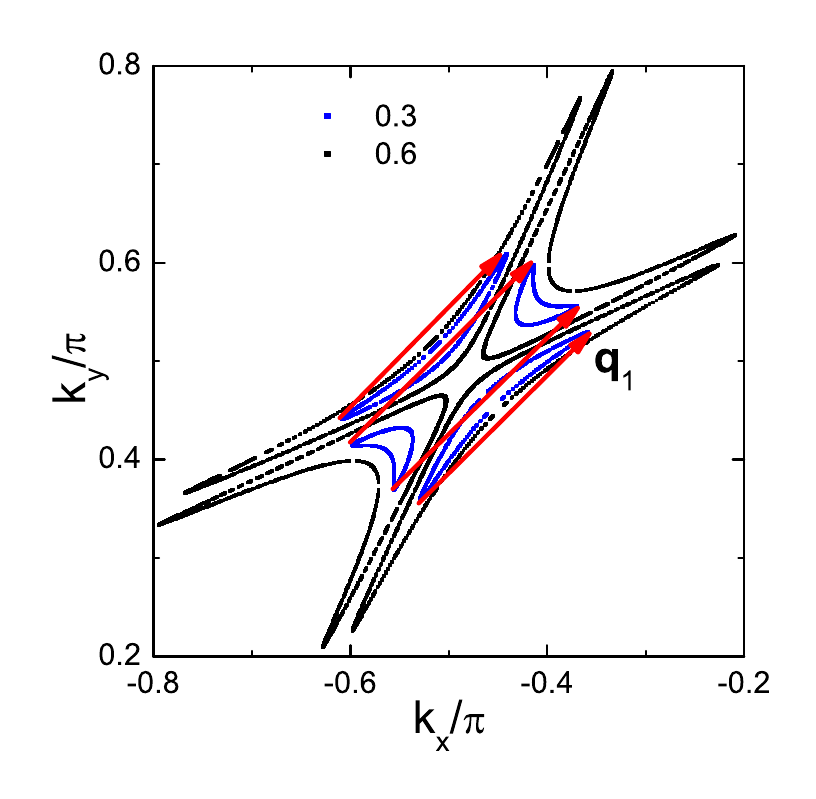}
 \caption{\label{0.3&0.6} (Color online) Similar to Fig. \ref{0.2}(b), but at $|\omega|/\Delta_{0}=0.3$ (blue) and $0.6$ (black).}
\end{figure}

\begin{figure}
\includegraphics[width=1\linewidth]{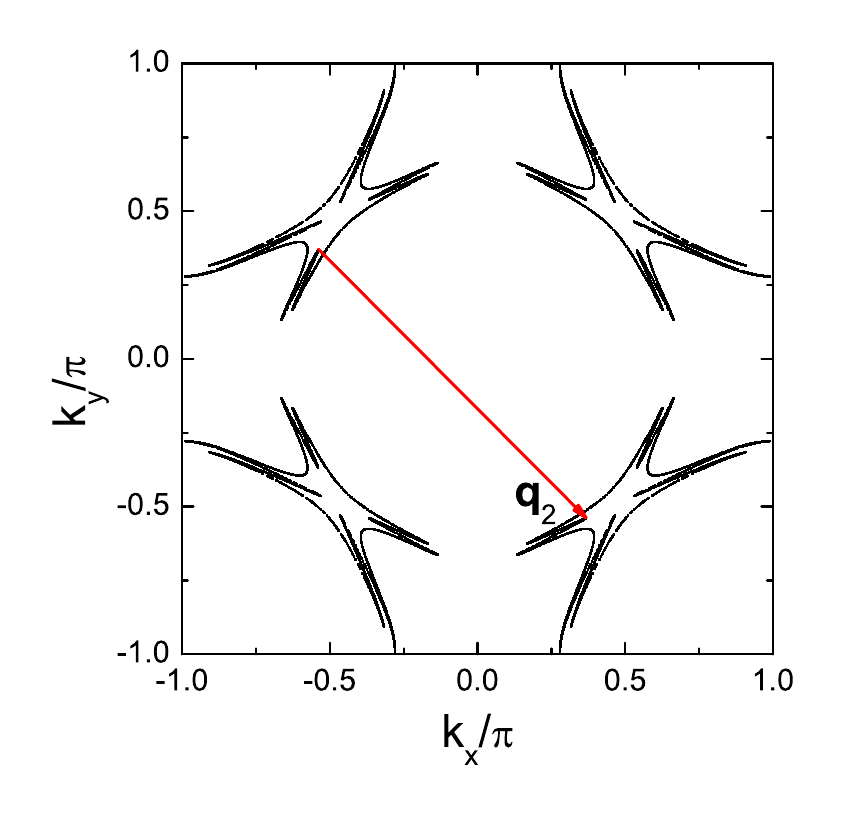}
 \caption{\label{0.7} (Color online) Similar to Fig. \ref{0.2}(a), but at $|\omega|/\Delta_{0}=0.7$.}
\end{figure}

\begin{figure*}
\includegraphics[width=1\linewidth]{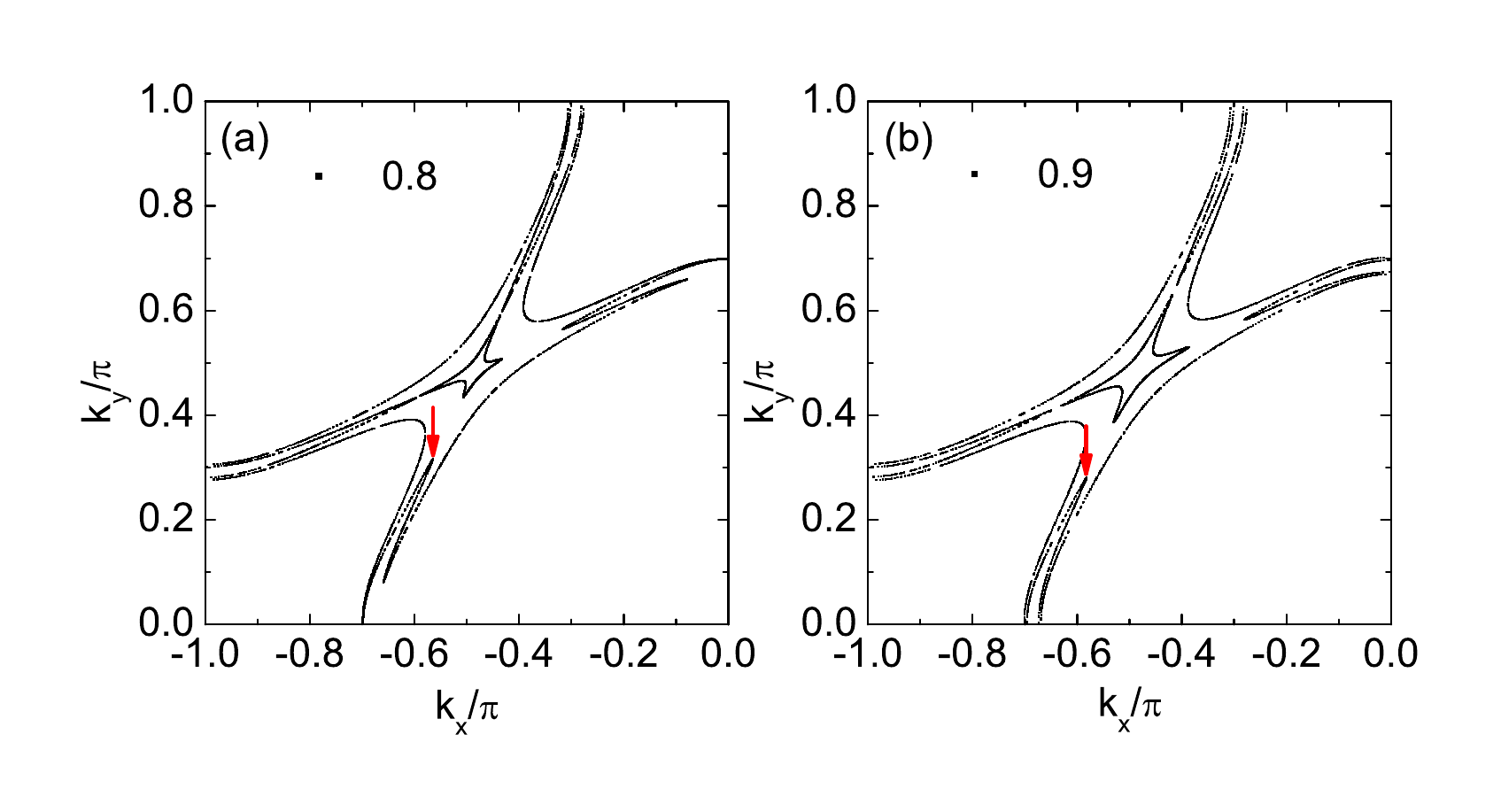}
 \caption{\label{0.8&0.9} (Color online) The CECs at $|\omega|/\Delta_{0}=0.8$ (a) and $0.9$ (b). Here we show only the CECs in the second quadrant. The red arrows denote one of the CEC tips connected by $\mathbf{q}_{2}$ as in Fig. \ref{0.7}.}
\end{figure*}

\begin{figure*}
\includegraphics[width=0.95\linewidth]{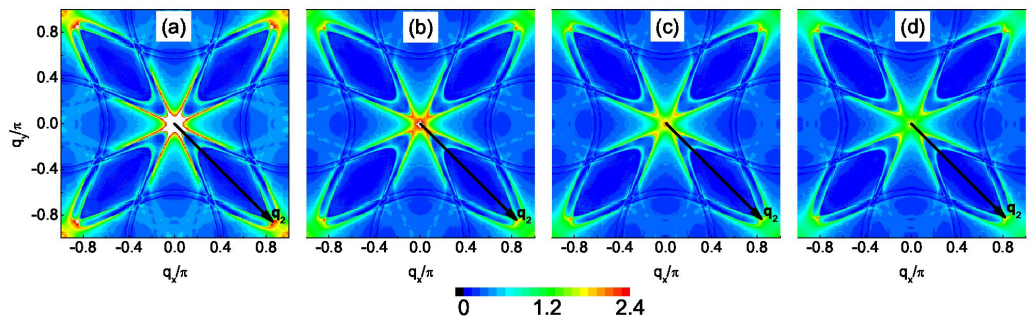}
 \caption{\label{-0.9to-1.2} (Color online) $|\rho(\mathbf{q},\omega)|$ at $\omega/\Delta_{0}=-0.9$ (a), $-1.0$ (b), $-1.1$ (c) and $-1.2$ (d).}
\end{figure*}

\begin{figure*}
\includegraphics[width=1\linewidth]{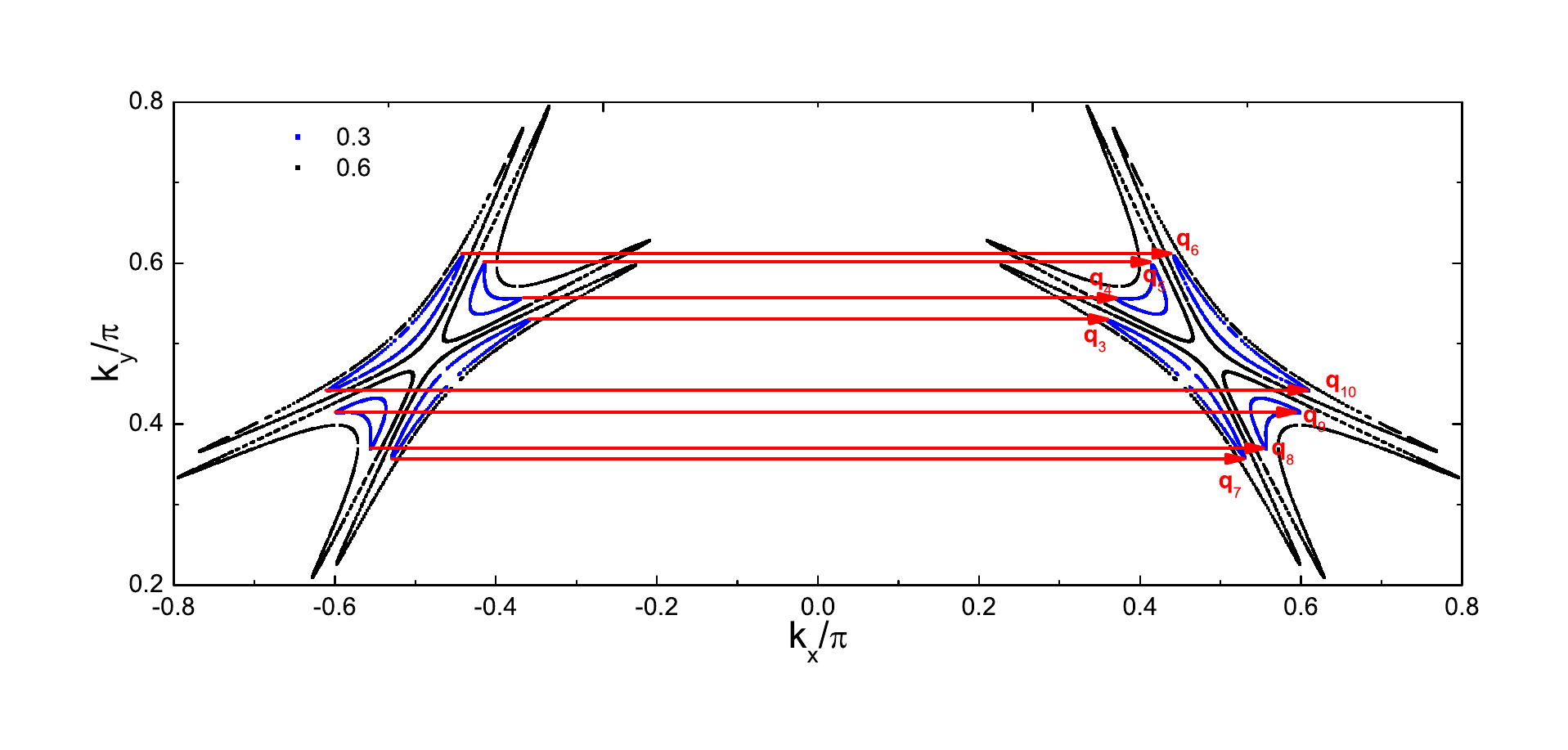}
 \caption{\label{other_wave_vectors} (Color online) The CECs at $|\omega|/\Delta_{0}=0.3$ (blue) and $0.6$ (black). Here we show the CECs in the first and second quadrants. $\mathbf{q}_{3},\ldots,\mathbf{q}_{10}$ are expected QPI wave vectors along the $(1,0)$ direction.}
\end{figure*}

\begin{figure*}
      \includegraphics[width=0.95\linewidth]{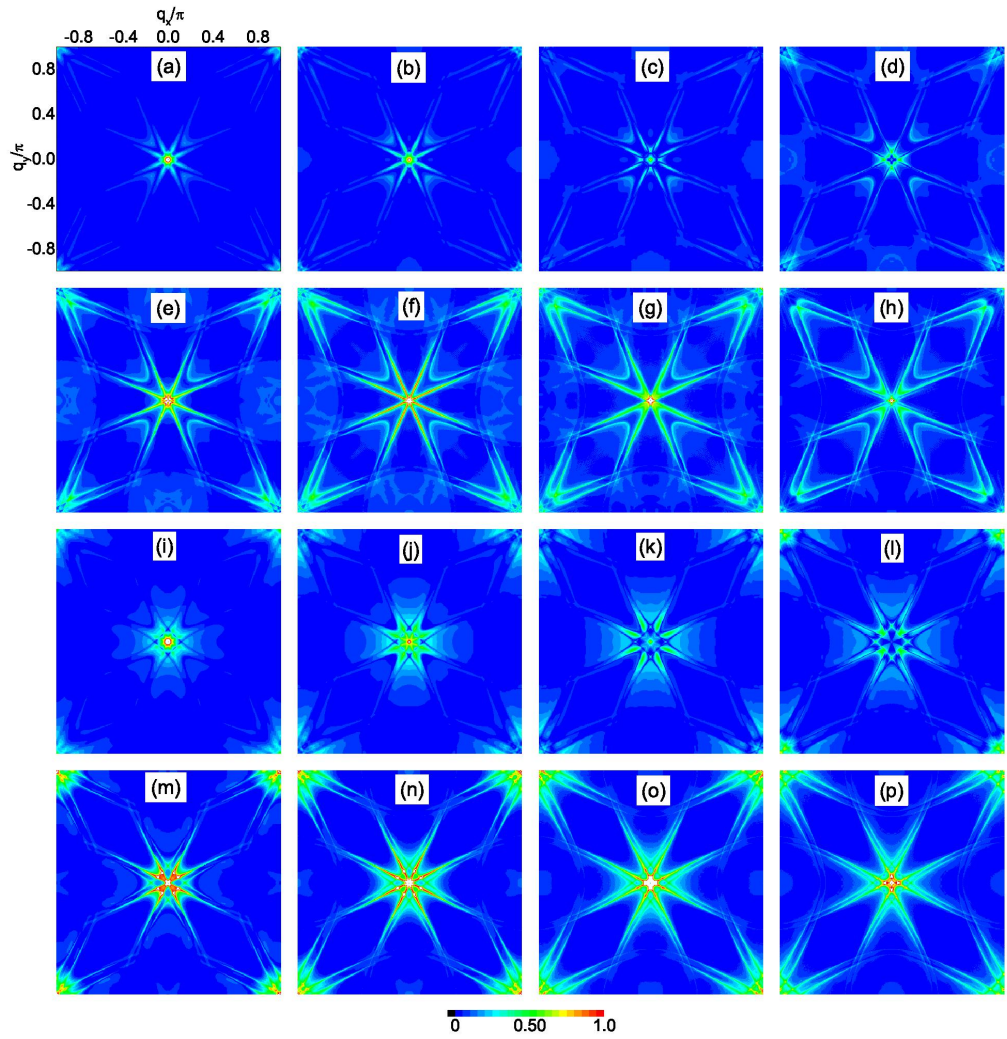}
\caption{\label{qpi_mag} (Color online) The same as Fig. \ref{qpi}, but for a magnetic impurity.}
\end{figure*}

\emph{Results}.---In the absence of the impurity, $\rho(\mathbf{r},\omega)$ is homogeneous in real space and we plot it as a function of the reduced energy $\omega/\Delta_{0}$ in Fig. \ref{dos}. As we can see, two pairs of SC resonance peaks are located at $\omega/\Delta_{0}\approx\pm0.77$ (denoted by the red arrows) and $\pm0.63$ (denoted by the black arrows), indicating the multi-gap structure resulting from the interplay of the complex band structure and pairing function. In the vicinity of $\omega/\Delta_{0}=0$, the spectrum is $V$-shaped and disperses linearly, suggesting that the SC gaps are nodal at $x=0.55$, as stated in Ref.~\onlinecite{qhwang}.

In the presence of the impurity, we plot $|\rho(\mathbf{q},\omega)|$ in Fig. \ref{qpi} and two main scattering wave vectors $\mathbf{q}_{1}$ and $\mathbf{q}_{2}$ can be found evolving with energy. $\mathbf{q}_{1}$ is located along the $q_{x}=\pm q_{y}$ directions and moves away from the origin as $|\omega|$ increases, which is clearly visible from $|\omega|/\Delta_{0}=0.2$ to $0.7$, except for $\omega/\Delta_{0}=-0.5$ and $0.3$ [see Figs. \ref{qpi}(d) and \ref{qpi}(j)], where $\mathbf{q}_{1}$ is masked by the high-intensity spots around it. However, for example, if we plot Fig. \ref{qpi}(d) by using a different color scale, $\mathbf{q}_{1}$ can still be found, as shown in Fig. \ref{-0.5}. At $|\omega|/\Delta_{0}=0.8$ and $0.9$, the intensity at $\mathbf{q}_{1}$ is too weak to be indentified. On the other hand, for $\mathbf{q}_{2}$, it is also located along the $q_{x}=\pm q_{y}$ directions and is visible only for $\omega/\Delta_{0}\leqslant-0.7$. As $\omega/\Delta_{0}$ varies from $-0.7$ to $-0.9$, $\mathbf{q}_{2}$ moves towards the origin, which can be seen in Figs. \ref{qpi}(f), \ref{qpi}(g) and \ref{qpi}(h).

Then we explain the $\omega$ dependence of $\mathbf{q}$ based on the evolution of the CEC. At $|\omega|/\Delta_{0}=0.2$, the banana-shaped CEC is shown in Fig. \ref{0.2} and $\mathbf{q}_{1}$ corresponds to the intrapocket scattering wave vector connecting the tips of the CEC [denoted by the red arrows in Fig. \ref{0.2}(b)]. From $|\omega|/\Delta_{0}=0.3$ to $0.6$, the CECs share the same topology and are shown in Fig. \ref{0.3&0.6}. In this case, $\mathbf{q}_{1}$ corresponds to not only the intrapocket, but also interpockdet scattering wave vectors as shown by the red arrows in Fig. \ref{0.3&0.6}. For example, from Figs. \ref{qpi}(e) and \ref{qpi}(m) we can see, at $\omega/\Delta_{0}=\pm0.6$, $q_{1x}=q_{1y}\approx(0.41\pm0.04)\pi$. At the same $\omega$, the CEC tips in Fig. \ref{0.3&0.6} are located at $(k_{x}/\pi,k_{y}/\pi)=(-0.6,0.23)/(-0.23,0.6)$, $(-0.63,0.21)/(-0.21,0.63)$, $(-0.79,0.34)/(-0.34,0.79)$ and $(-0.77,0.37)/(-0.37,0.77)$. The derived $\mathbf{q}_{1}/\pi$ should be $(0.37,0.37)$, $(0.42,0.42)$, $(0.45,0.45)$ and $(0.4,0.4)$, agrees fairly well with those in Figs. \ref{qpi}(e) and \ref{qpi}(m). With increasing $|\omega|$, the size of the banana increases, leading to an increased $|\mathbf{q}_{1}|$. As $|\omega|/\Delta_{0}$ varies to $0.7$, the CEC topology changes and is shown in Fig. \ref{0.7}. In this case, $\mathbf{q}_{2}$ emerges due to the appearance of additional tips of the CEC. In Fig. \ref{0.7}, the tips connected by $\mathbf{q}_{2}$ are located at $(k_{x}/\pi,k_{y}/\pi)=(-0.54,0.37)/(0.37,-0.54)$ and the derived $\mathbf{q}_{2}/\pi\approx(0.91,-0.91)$, consistent with the $\mathbf{q}_{2}$ shown in Fig. \ref{qpi}(f). Further increasing $|\omega|/\Delta_{0}$ to $0.8$ and $0.9$, the CECs are shown in Fig. \ref{0.8&0.9} and the topology again changes with an additional closed area in between. The CEC tips marked by the red arrows are now located at $(k_{x}/\pi,k_{y}/\pi)=(-0.56,0.32)$ in Fig. \ref{0.8&0.9}(a) and $(-0.58,0.28)$ in Fig. \ref{0.8&0.9}(b). Thus the derived $\mathbf{q}_{2}/\pi$ should be $(0.88,-0.88)$ and $(0.86,-0.86)$ for $|\omega|/\Delta_{0}=0.8$ and $0.9$, respectively, both of which are consistent with those shown in Figs. \ref{qpi}(g) and \ref{qpi}(h). Furthermore, at $|\omega|/\Delta_{0}=1$, $1.1$ and $1.2$, the CEC topologies are the same as those for $|\omega|/\Delta_{0}=0.8$ and $0.9$, while the tips move to $(k_{x}/\pi,k_{y}/\pi)=(-0.6,0.25)$, $(-0.61,0.22)$ and $(-0.62,0.2)$ for $|\omega|/\Delta_{0}=1$, $1.1$ and $1.2$, respectively. The derived $\mathbf{q}_{2}/\pi$ becomes to be $(0.85,-0.85)$, $(0.83,-0.83)$ and $(0.82,-0.82)$ for $|\omega|/\Delta_{0}=1$, $1.1$ and $1.2$, respectively and we can see from Fig. \ref{-0.9to-1.2}, this is indeed the case. Thus we conclude that $\mathbf{q}_{1}$ and $\mathbf{q}_{2}$ indeed correspond to those scattering wave vectors shown in Figs. \ref{0.3&0.6} and \ref{0.7} while their evolution closely follows the evolution of the CEC.

In addition, apart from the above mentioned $\mathbf{q}_{1}$ and $\mathbf{q}_{2}$, naively we would expect that more QPI wave vectors connecting other tips of the CEC should exist. For example, from Fig. \ref{other_wave_vectors} we can see, the CECs share the same topology when $|\omega|/\Delta_{0}$ varies from $0.3$ to $0.6$ and along the $(1,0)$ direction, there should exist QPI wave vectors like $\mathbf{q}_{3},\ldots,\mathbf{q}_{10}$. At $|\omega|/\Delta_{0}=0.3$, $\mathbf{q}_{3}/\pi=(0.72,0)$, $\mathbf{q}_{4}/\pi=(0.74,0)$, $\mathbf{q}_{5}/\pi=(0.83,0)$, $\mathbf{q}_{6}/\pi=(0.88,0)$, $\mathbf{q}_{7}/\pi=(0.94,0)$, $\mathbf{q}_{8}/\pi=(0.89,0)$, $\mathbf{q}_{9}/\pi=(0.8,0)$, $\mathbf{q}_{10}/\pi=(0.79,0)$ ($\mathbf{q}_{7}$ to $\mathbf{q}_{10}$ are umklapp processes). As $|\omega|/\Delta_{0}$ increases to $0.6$, $\mathbf{q}_{3},\ldots,\mathbf{q}_{10}$ move towards the origin and are located at $(0.45,0)$, $(0.42,0)$, $(0.66,0)$, $(0.74,0)$, $(0.8,0)$, $(0.74,0)$, $(0.41,0)$ and $(0.46,0)$, respectively. However in Fig. \ref{qpi} we cannot find such wave vectors evolving with energy. In order to verify this, we repeated the above calculations for $V_{s}=8\Delta_{0}$ and the results remain qualitatively the same. The reason may be: as we know, in single-band $d$-wave superconductors, the actual intensity of the scattering wave vectors depends on the SC coherence factors and will be either suppressed or enhanced when changing the nonmagnetic impurity to a magnetic one.~\cite{QPI} In contrast, in multi-orbital superconductors, the intensity of the scattering wave vectors is determined by not only the SC coherence factors, but also the matrix elements of the unitary transformation between the orbital and band bases. In the present model for BiS$_{2}$-based superconductors, it is very likely that the matrix elements associated with the above mentioned scattering processes ($\mathbf{q}_{3},\ldots,\mathbf{q}_{10}$) are vanishingly small, thus those QPI wave vectors cannot be detected experimentally.

Furthermore, we also studied the magnetic impurity case simply by rewriting Eq. (\ref{u}) as
\begin{eqnarray}
\label{u_mag}
U_{lm}=\begin{cases}
V_{s}&\text{$l=m=1, 2, 7, 8$},\\
-V_{s}&\text{$l=m=3, 4, 5, 6$},\\
0&\text{otherwise}.
\end{cases}
\end{eqnarray}
The obtained QPI spectra are shown in Fig. \ref{qpi_mag}. We found that for magnetic impurity scattering, the intensity of the QPI patterns is weaker than the nonmagnetic impurity case and most importantly, there are no clear and sharp QPI vectors evolving with energy. While the disappearance of $\mathbf{q}_{1}$ and $\mathbf{q}_{2}$ may be attributed to the effects of the SC coherence factors, the lack of significant QPI vectors along the $(1,0)$ direction for both the nonmagnetic and magnetic scatterings is further proved to be due to the matrix elements of the unitary transformation between the orbital and band bases.

\emph{Summary}.---In summary, we have studied the QPI spectra in BiS$_{2}$-based superconductors, by taking into account the SOC. The pairing symmetry is assumed to be the $d^{*}_{x^{2}-y^{2}}$-wave as proposed in Ref. \onlinecite{qhwang} and we are seeking for the experimental evidence to test it. Specifically, we found two distinct scattering wave vectors evolving with energy and their evolution closely follows the evolution of the CEC. Thus the QPI spectra presented in this paper can be compared with future STM experiments to test whether the pairing symmetry is $d^{*}_{x^{2}-y^{2}}$-wave for BiS$_{2}$-based superconductors. Furthermore, we notice that there exist multiple tips of the CEC, especially at higher $|\omega|$ and there should exist other wave vectors connecting these tips. However, they did not show up in the QPI spectra, possibly due to the effects of the matrix elements of the unitary transformation between the orbital and band bases.

We thank Y. Xiong and Y. Yang for helpful discussions. This work was supported by NSFC (Grants No. 11204138 , No. 11175087, No. 11374005 and No. 11023002), the Ministry of Science and Technology of China (Grants No. 2011CBA00108 and 2011CB922101), NSF of Jiangsu Province of China (Grant No. BK2012450), NSF of Shanghai(Grant No.13ZR1415400), Program of Natural Science Research of Jiangsu Higher Education Institutions of China (Grant No. 12KJB140009), SRFDP (Grant No. 20123207120005), China Postdoctoral Science Foundation (Grant No. 2012M511297) and NCET (Grant No. NCET-12-0626). The numerical calculations in this paper have been done on the IBM Blade cluster system in the High Performance Computing Center (HPCC) of Nanjing University.

\end{document}